\documentclass[12pt,twoside,a4paper]{article} 
\usepackage{amsmath,amssymb,latexsym,theorem,bbm,shapepar,natbib,epsfig}
\newfont{\msa}{msam10 scaled\magstep1}
\newfont{\ssmsa}{msam9}

\def\crps{\mathop{\hbox{\rm crps}}}

\numberwithin{equation}{section}

\title{Probabilistic wind speed forecasting in Hungary}

\author{{\sc S\'andor Baran} and {\sc D\'ora Nemoda}\\ 
         Faculty of Informatics, University of Debrecen\\
         Kassai \'ut 26, H--4028 Debrecen, Hungary \\ [5mm]
       {\sc Andr\'as Hor\'anyi} \\   
       Hungarian Meteorological Service \\
       P.O. Box 38, H--1525 Budapest, Hungary     
     }

\date{}

\begin{document}
\pagestyle{myheadings}

\maketitle

\begin{abstract}
Prediction of various weather quantities is mostly based on
deterministic numerical weather forecasting models.  Multiple runs of
these models with different initial conditions result ensembles of
forecasts which are applied for estimating the distribution of
future weather quantities. However, the ensembles are usually
under-dispersive and uncalibrated, so post-processing is required.

In the present work Bayesian Model Averaging (BMA) is applied for
calibrating ensembles of wind speed forecasts produced by the
operational Limited Area Model Ensemble Prediction System of the
Hungarian Meteorological Service (HMS). 

We describe two possible BMA models for wind speed data of the HMS and
show that BMA post-processing significantly improves the calibration
and precision of forecasts.

\end{abstract}

\section{Introduction}
   \label{sec:sec1}

The aim of weather forecasting is to give a good prediction of the
future states of the atmosphere on the basis of present observations
and mathematical models describing the dynamics (physical behaviour)
of the atmosphere.    
These models consist of sets of non-linear partial differential
equations which have only numerical solutions. The problem with these
numerical weather prediction models is that the solutions highly
depend on the initial conditions which are always in a way or in
another not fully accurate. A
possible solution to address this problem is to run the model with different
initial conditions and produce ensembles of forecasts. With the help
of ensembles one can estimate the distribution of future weather
variables which leads us to probabilistic weather forecasting \citep{gr}.
The ensemble
prediction method was proposed by \citet{leith} and since its first
operational implementation \citep{btmp,tk} it became a widely used
technique all over the world. However, despite e.g. the ensemble
mean gives a better estimate of a meteorological quantity than most or
all of the ensemble members, the ensemble is usually under-dispersive
and in this way, uncalibrated. This phenomena was observed at several
operational ensemble prediction systems, for an overview see
e.g. \citet{bhtp}. 

The Bayesian model averaging (BMA) method for
post-processing ensembles in order to calibrate them was introduced by
\citet{rgbp}. The basic idea of BMA is that to each ensemble member
forecast corresponds a conditional probability
density function (PDF) that can be interpreted as the
conditional PDF of the future weather quantity provided the considered
forecast is the best one. Then the BMA predictive PDF of the future
weather quantity is the weighted sum of the individual PDFs
corresponding to the ensemble members and the weights are based on the
relative performances of the ensemble members during a given training period.
In \citet{rgbp} the BMA
method was successfully applied to obtain 48 hour forecasts of surface
temperature and sea level pressure in the North American Pacific
Northwest based on the 5 members of the University of Washington
Mesoscale Ensemble \citep{gm}. These weather quantities can be
modeled by normal distributions, so the predictive PDF is a Gaussian mixture.
Later \citet{srgf} developed a discrete-continuous BMA model for precipitation
forecasting, where the discrete part corresponds to the event of no
precipitation, while the cubic root of the precipitation amount
(if it is positive) is modeled by a gamma distribution. In
\citet{sgr10} the BMA method was used for wind speed forecasting and 
the component PDFs follow gamma distribution. Finally, using von Mises
distribution to model angular data 
\citet{bgrgg} introduced a BMA scheme to predict surface wind
direction.

In the present work we apply the BMA method for calibrating ensemble
forecasts of wind speed produced by the operational Limited
Area Model Ensemble  
Prediction System (LAMEPS) of the Hungarian Meteorological Service
(HMS) called ALADIN-HUNEPS \citep{hagel, horanyi}. ALADIN-HUNEPS
covers a large part of Continental Europe with a horizontal resolution
of 12 km and it is obtained by dynamical downscaling (by the ALADIN
limited area model) of the global
ARPEGE based PEARP system of M\'et\'eo France \citep{hkkr,dljn}. The
ensemble consists of 11 members, 10 initialized from perturbed initial
conditions and one control member from the unperturbed analysis. This
construction implies that the ensemble contains groups of exchangeable
forecasts (the ensemble members cannot be distinguished), so for
post-processing one has to use the modification of 
BMA as suggested by \citet{frg}.

\section{Data}
  \label{sec:sec2}

As it was mentioned in the Introduction, BMA post-processing
of ensemble predictions was applied for wind speed data obtained from
the HMS. The 
data file contains 11 member ensembles (10 forecasts started from perturbed
initial conditions and one control) of 42 hour 
forecasts for 10 meter wind speed (given in m/s)
for 10 major cities in 
Hungary (Miskolc, Szombathely, Gy\H or, Budapest, Debrecen, Ny\'\i regyh\'aza,
Nagykanizsa, P\'ecs, Kecskem\'et, Szeged) produced by the
ALADIN-HUNEPS system of the HMS, together with the corresponding
validating observations, for the period between October 1, 2010 and March 25,
2011. The forecasts are initialized at 18 UTC, the startup speed of
the anemometers measuring the validating observations is $0.1$
m/s. The data set is fairly complete, since there are only two days
(18.10.2010 and 15.02.2011) when three ensemble members are missing
for all sites and one day (20.11.2010) when no forecasts are available. 

\begin{figure}[t]
\begin{center}
\leavevmode
\epsfig{file=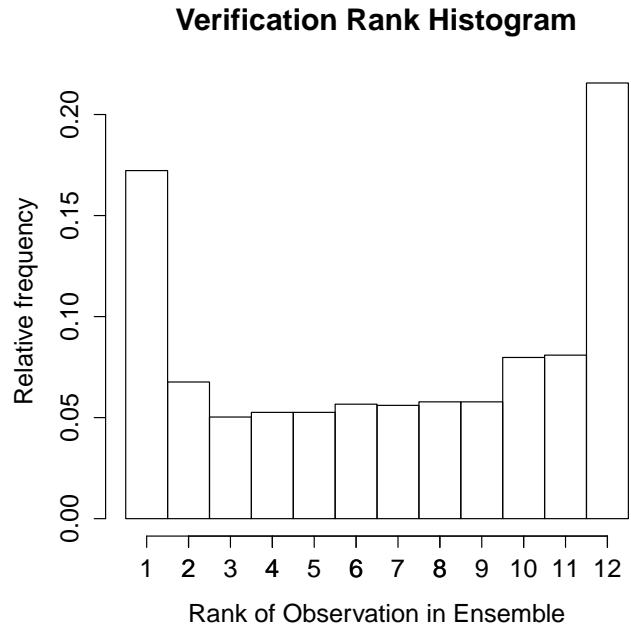,height=9cm, angle=-90}
\caption{Verification rank histogram of the 11-member ALADIN-HUNEPS
  ensemble. Period: October 1, 2010 -- March 25, 2011.} 
\label{fig:fig1}
\end{center}
\end{figure}
Figure \ref{fig:fig1} shows the verification rank histogram of the raw
ensemble, that is the histogram of ranks of validating
observations with respect to the corresponding ensemble
forecasts. This histogram is 
far from the desired uniform distribution, in most of the cases the
ensemble members either underestimate, or overestimate the validating
observations (the ensemble range contains the observed wind speed only
in $61.21\%$ of the cases). Hence, the ensemble is under-dispersive
and in this way it is uncalibrated.

\section{The model and diagnostics}
   \label{sec:sec3}

To obtain a probabilistic forecast of wind speed  the modification of 
BMA gamma model of \citet{sgr10} for ensembles with
exchangeable members \citep{frg} was used. The first idea is to have two
exchangeable groups: one contains 
the control denoted by \ $f_c$, \ the other one the 10 ensemble members
corresponding to the different perturbed initial conditions which are denoted by
\ $f_{\ell,1},\ldots ,f_{\ell,10}$, \ respectively. \ In this way we assume
that the probability density function (PDF) of the forecasted
wind speed \ $x$ \ equals:    
\begin{align}
  \label{eq:eq3.1}
p(x | f_c,f_{\ell,1},\ldots ,
f_{\ell,10};b_0,b_1,c_0,c_1)=&\, \omega
g(x;f_c,b_0,b_1,c_0,c_1) \\ &+  \frac 
{1-\omega}{10} \sum_{j=1}^{10}  g(x;f_{\ell,j},b_0,b_1,c_0,c_1), \nonumber 
\end{align} 
where \ $\omega\in [0,1]$, \ and \  $g$ \
is the conditional PDF corresponding to the ensemble members.
As we are working with wind speed data, 
\ $g(x;f,b_0,b_1,c_0,c_1)$ \ is a gamma PDF with mean
\ $b_0+b_1 f$ \ and standard deviation \ $c_0+c_1 f$. \ Here we restrict both
the mean and the standard deviation parameters to constant values for all
ensemble members, which reduces the number of parameters and
simplifies calculations. 
Mean parameters \ $b_0,\ b_1$ \ are estimated with
the help of linear regression, while weight  \ $\omega$ \ and standard deviation
parameters \ $c_0, \ c_1$, \ by maximum likelihood method, using training
data consisting of ensemble members and verifying observations from
the preceding \ $n$ \ days (training period). In order to handle the problem
that the wind speed values under 0.1 m/s are considered to be zero, the
maximum likelihood (ML) method for gamma distributions suggested by
\citet{wilks1} is 
applied, while the maximum of the likelihood
function is found with the help of EM algorithm \citep{mclk}. For more
details see \citet{sgr10,frg}. Once the estimated 
parameters for a given day are available,  one can
use either the mean or the median of the predictive PDF
\eqref{eq:eq3.1} as a point forecast.

\begin{figure}[t]
\begin{center}
\leavevmode
\epsfig{file=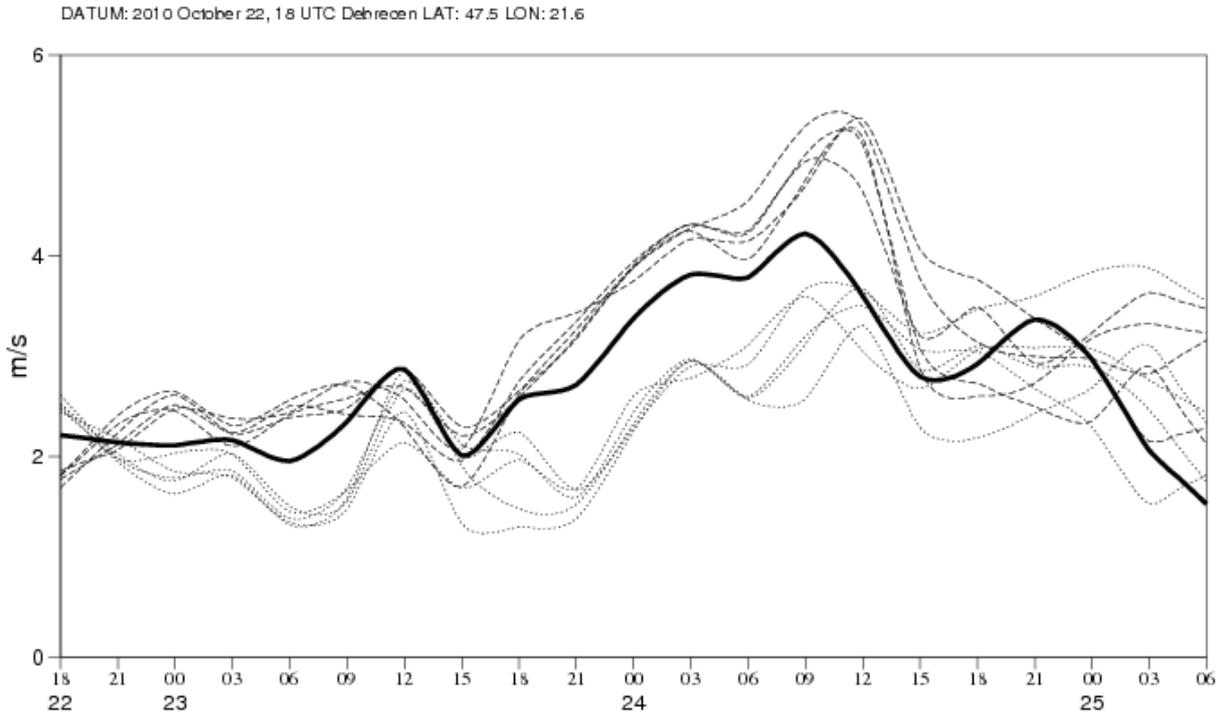,height=10cm}
\caption{Plume diagram of ensemble forecast of 10 meter wind speed
  for Debrecen initialized at 18 UTC, 22.10.2010.} 
\label{fig:fig2}
\end{center}
\end{figure}
Based on a more careful look on the ensemble members there are some
differences in the generation of the ten 
exchangeable ensemble members. To obtain them only five perturbations
are calculated and then they are added to (odd numbered members) and
subtracted from (even numbered members) the unperturbed initial
conditions \citep{horanyi}. Figure \ref{fig:fig2} shows the plume
diagram of ensemble forecast of 10 meter wind speed
for Debrecen initialized at 18 UTC, 22.10.2010. (solid line: control;
dotted line: odd numbered members, dashed line: even numbered
members). This diagram clearly illustrates that the behaviour of
ensemble member groups \ $\{f_{\ell,1}, \ 
f_{\ell,3}, \ f_{\ell,5}, \ f_{\ell,7}, \ f_{\ell,9}\}$ \ and \ $\{f_{\ell,2}, \
f_{\ell,4}, \ f_{\ell,6}, \ f_{\ell,8}, \ f_{\ell,10}\}$ \ really differ
from each other. Therefore, in this way one can also consider a model with three
exchangeable groups: control, odd numbered exchangeable members and even
numbered exchangeable members. This idea leads to the following PDF
of the forecasted wind speed \ $x$:
\begin{align}
  \label{eq:eq3.2}
q(x | f_c,f_{\ell,1},\ldots ,
f_{\ell,10};&\,b_0,b_1,c_0,c_1)= \omega_c
g(x;f_c,b_0,b_1,c_0,c_1) \\ &+ 
\sum_{j=1}^{5} \big(\omega_o g(x;f_{\ell,2j-1},b_0,b_1,c_0,c_1)+ 
\omega_e g(x;f_{\ell,2j},b_0,b_1,c_0,c_1)\big), \nonumber 
\end{align} 
where for weights \ $\omega_c,\omega_o,\omega_e\in[0,1]$ \ we have  \
$\omega_c+5\omega_o+5\omega_e=1$, \ while PDF \ $g$ \ and parameters \
$b_0,b_1,c_0,c_1$ \ are the same as for the model
\eqref{eq:eq3.1}. Obviously, both the weights and the parameters 
can be estimated in the same way as before.

\begin{figure}[t!]
\begin{center}
\leavevmode
\hbox{
\epsfig{file=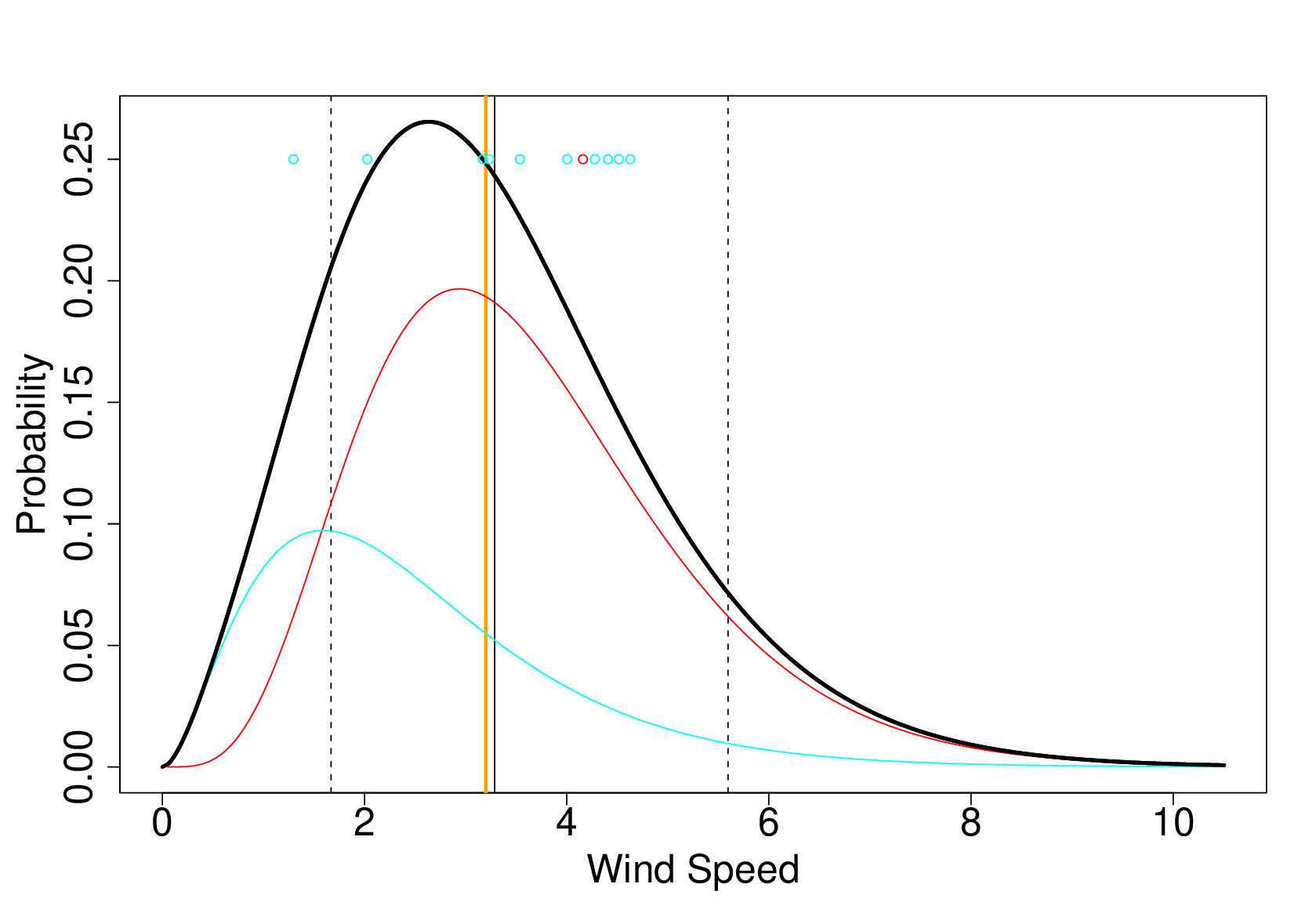,height=8cm, angle=-90} \quad
\epsfig{file=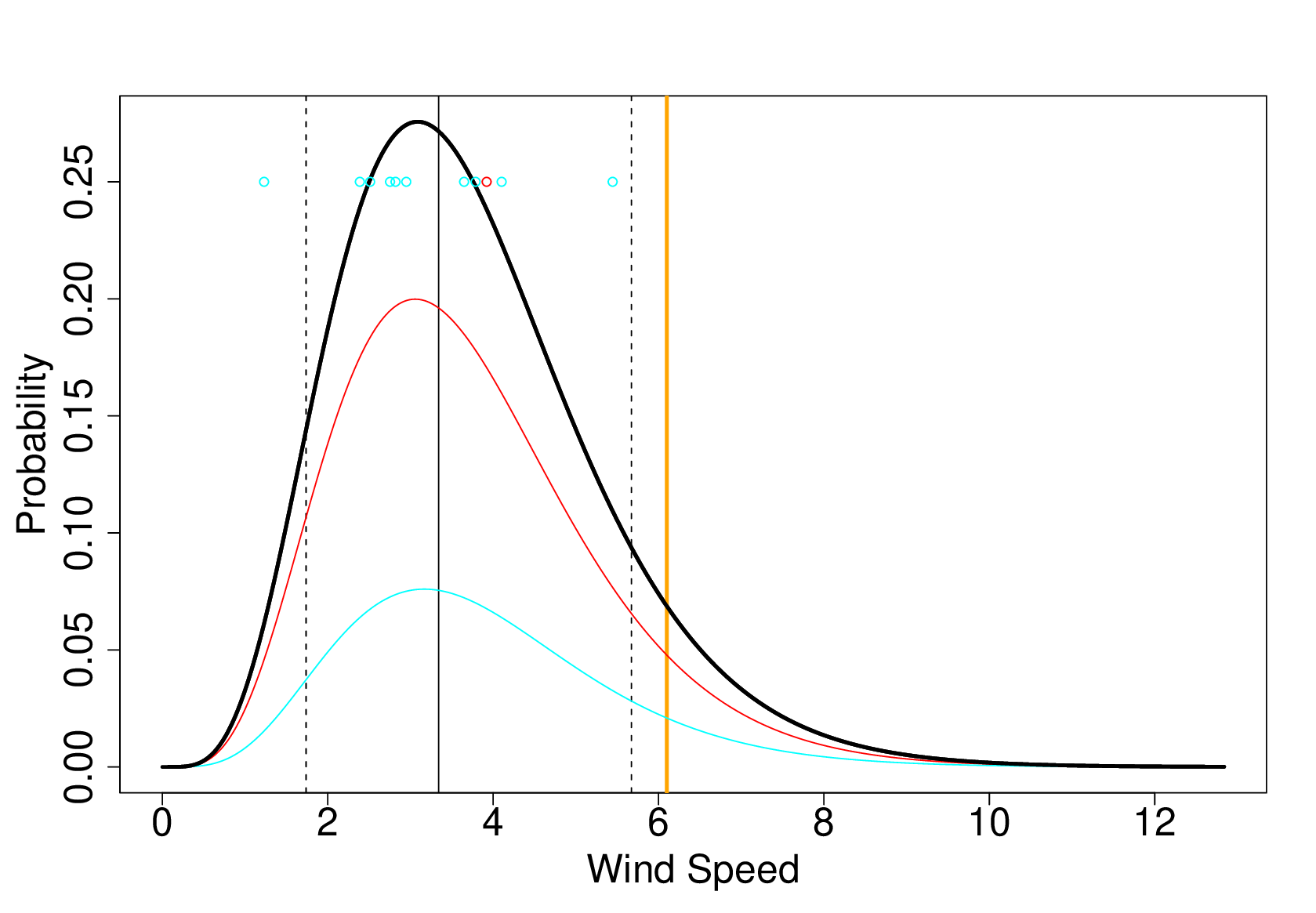,height=8cm, angle=-90}}

\centerline{\hbox to 9 truecm {\scriptsize (a) \hfill (b)}}

\hbox{
\epsfig{file=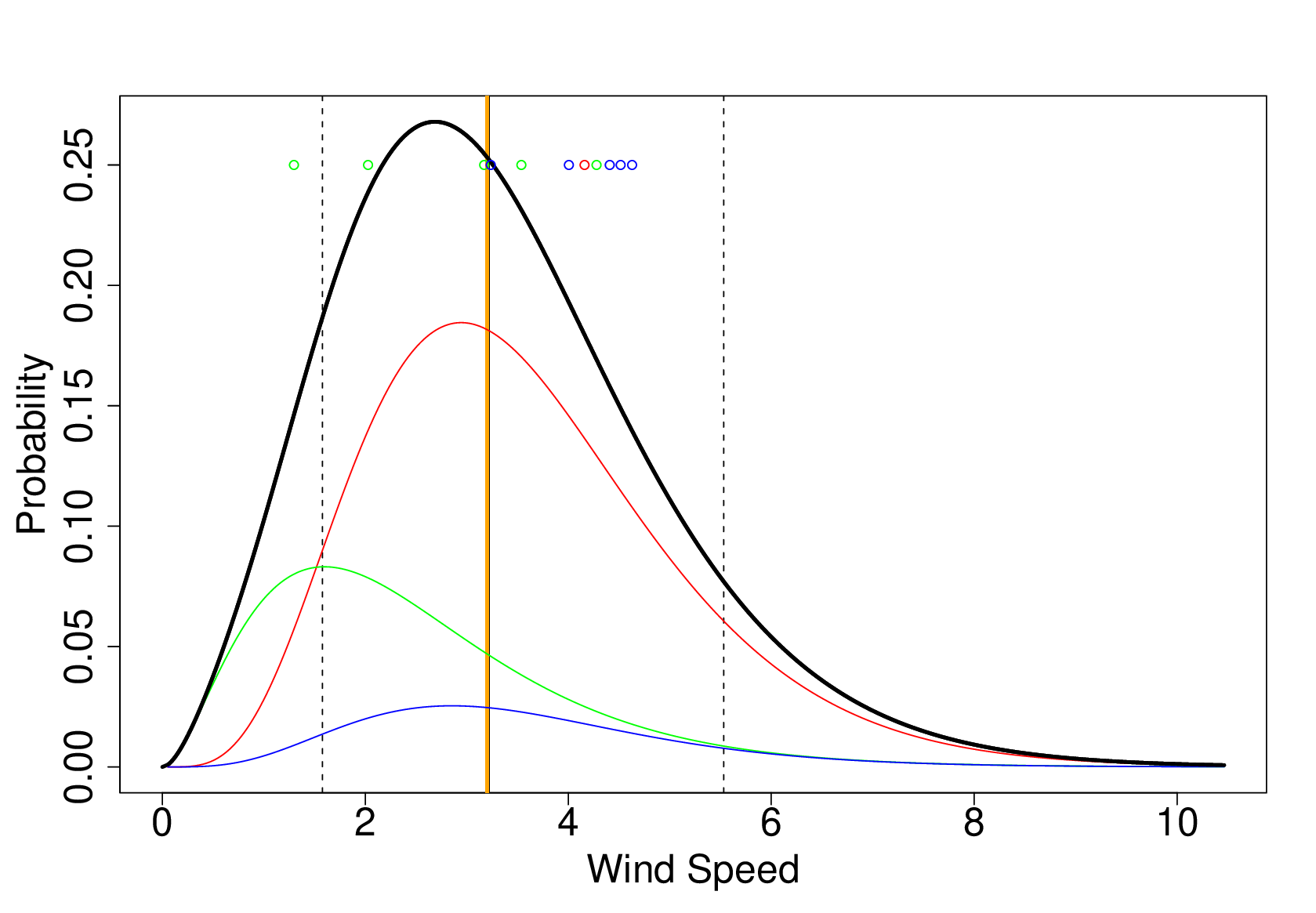,height=8cm, angle=-90} \quad
\epsfig{file=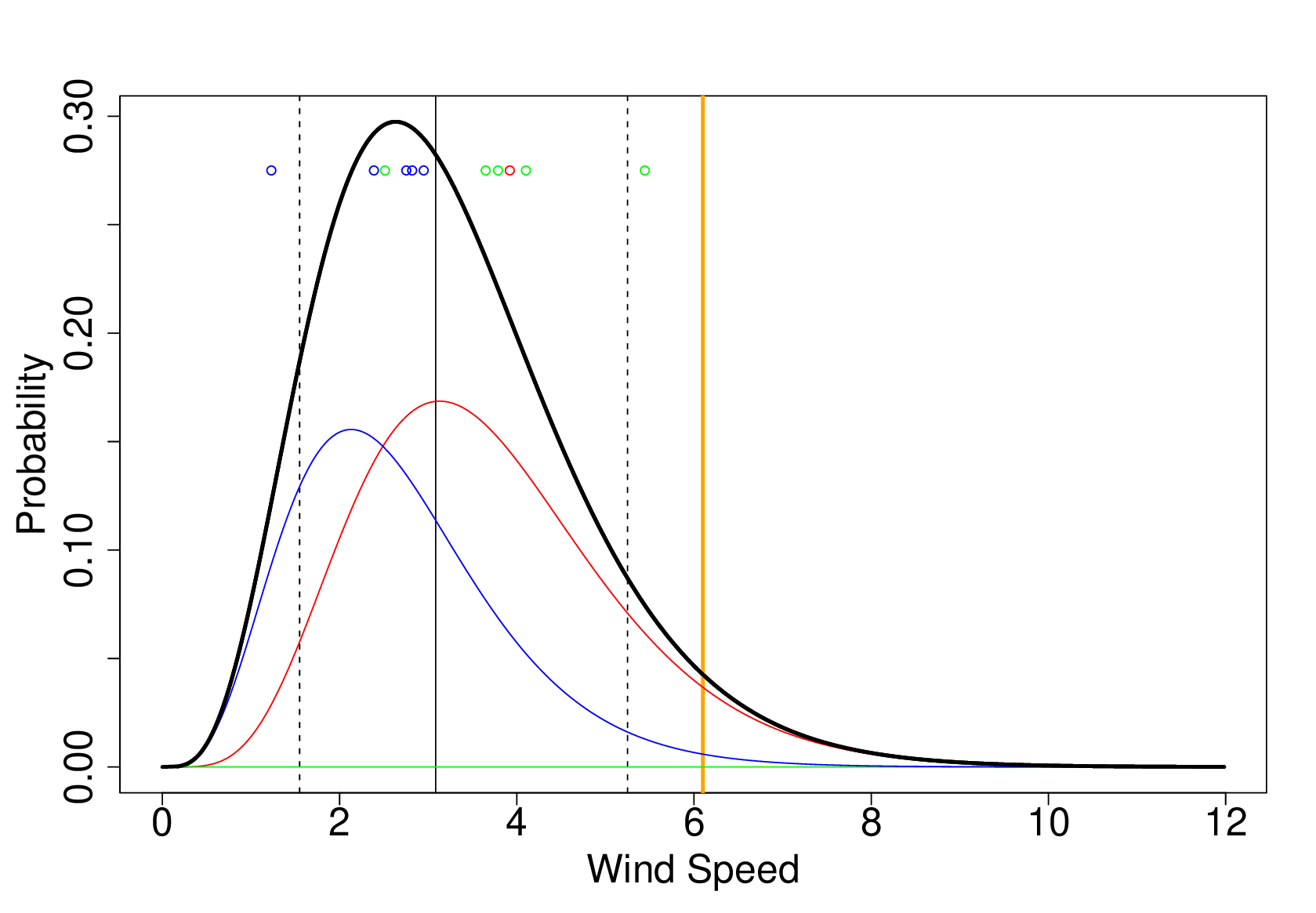,height=8cm, angle=-90}}

\centerline{\hbox to 9 truecm {\scriptsize (c) \hfill (d)}}
\caption{Ensemble BMA PDFs (overall: thick black line; control: red
  line; sum of exchangeable members on (a) and (b): light blue line;
  on (c) and (d): green (odd members) and blue (even members) lines),
  ensemble members (circles with the same colours as the corresponding
  PDFs), ensemble BMA median forecasts (vertical black line),
  verifying observations (vertical orange line) and the first and last
deciles (vertical dashed lines) for wind speed in Debrecen for
  models  \eqref{eq:eq3.1}: (a) 30.12.2010, (b) 17.03.2011; and
  \eqref{eq:eq3.2}: (c) 30.12.2010, (d) 17.03.2011.} 
\label{fig:fig3}
\end{center}
\end{figure}
As an illustration we consider the data and forecasts for 
Debrecen for two
different dates 30.12.2010 and 17.03.2011 for models \eqref{eq:eq3.1}
and \eqref{eq:eq3.2}.
Figures \ref{fig:fig3}a and \ref{fig:fig3}b show the  PDFs of
the two groups in model \eqref{eq:eq3.1}, the overall 
PDFs, the median forecasts,
the verifying observations, the first and last
deciles  and the ensemble members. The same functions and
quantities can be seen on Figures \ref{fig:fig3}c and \ref{fig:fig3}d,
where besides the overall PDF we have three component PDFs and three
groups of ensemble members. On 30.12.2010 the spread of the ensemble
members is quite fair and the ensemble range contains the validating
observation (3.2 m/s). In this case the ensemble mean (3.5697 m/s)
overestimates, while BMA median forecasts corresponding to the two- and
three-group models (3.2876 m/s and 3.2194 m/s, respectively) are
pretty close to the true wind speed.  A different situation is
illustrated on Figures \ref{fig:fig3}b and \ref{fig:fig3}d, where the
spread of the ensemble is even higher, but all
ensemble members underestimate the validating observation (6.1
m/s). Obviously, the same holds for the ensemble mean (3.2323 m/s) and
due to the bias correction the BMA median forecasts corresponding to
models \eqref{eq:eq3.1} and 
\eqref{eq:eq3.2} also give bad results (3.3409 m/s and 3.0849 m/s,
respectively).

\begin{figure}[t!]
\begin{center}
\leavevmode
\epsfig{file=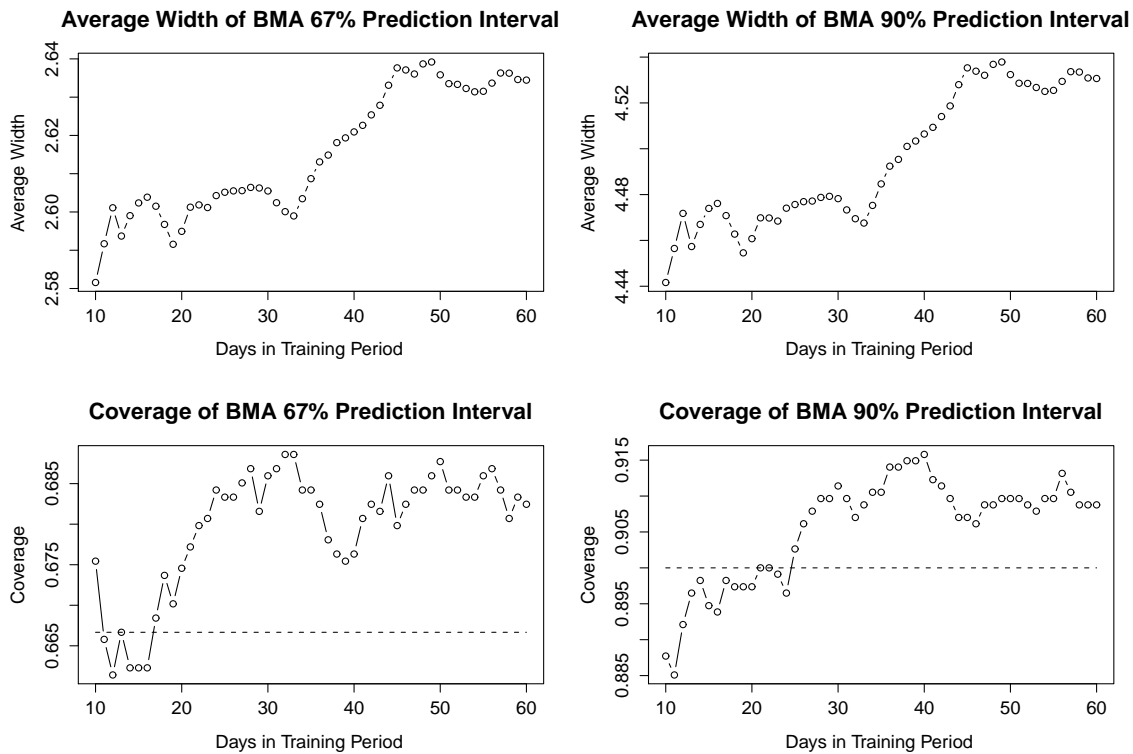,height=15cm, angle=-90}
\caption{Average widths and coverages of $66.7\,\%$ and $90\,\%$
 BMA  prediction intervals 
  corresponding to two-group model \eqref{eq:eq3.1} for various training period
  lengths.}   
\label{fig:fig4}
\end{center}
\end{figure}
To check the performance of probabilistic forecasts based on
models \eqref{eq:eq3.1} and \eqref{eq:eq3.2} and the
corresponding point forecasts, as a reference we use the ensemble mean
and the ensemble median. We compare the mean absolute errors (MAE) and
the root mean square errors (RMSE) of
these point forecasts and also the mean continuous ranked probability scores
(CRPS) \citep{wilks2,grjasa} and the coverages and average widths of
$66.7\,\%$ and $90\,\%$ prediction intervals of the BMA predictive probability
distributions and of the raw ensemble. We remark that for MAE and RMSE
the optimal point forecasts are the median and the mean, respectively
\citep{gneiting11, pinhag}.  Further, given a cumulative distribution
function (CDF) \ $F(y)$ \ and a real number \ $x$, \ the CRPS is defined as
\begin{equation*}
\crps\big(F,x\big):=\int_{-\infty}^{\infty}\big (F(y)-{\mathbbm 
  1}_{\{y \geq x\}}\big )^2{\mathrm d}y.
\end{equation*}
The mean CRPS of a probability forecast is the average of the CRPS values
of the predictive CDFs and corresponding validating observations taken
over all locations and time points considered. For the raw ensemble
the empirical CDF of the ensemble replaces the predictive CDF. The
coverage of a \ $(1-\alpha)100 \,\%, \ \alpha \in (0,1),$ \ prediction
interval is the proportion 
of validating observations located between the lower and upper \
$\alpha/2$ \ quantiles of the predictive distribution. For a
calibrated predictive PDF this value should be around \ $(1-\alpha)100
\,\%$.

\section{Results}
   \label{sec:sec4}

\begin{figure}[t!]
\begin{center}
\leavevmode
\epsfig{file=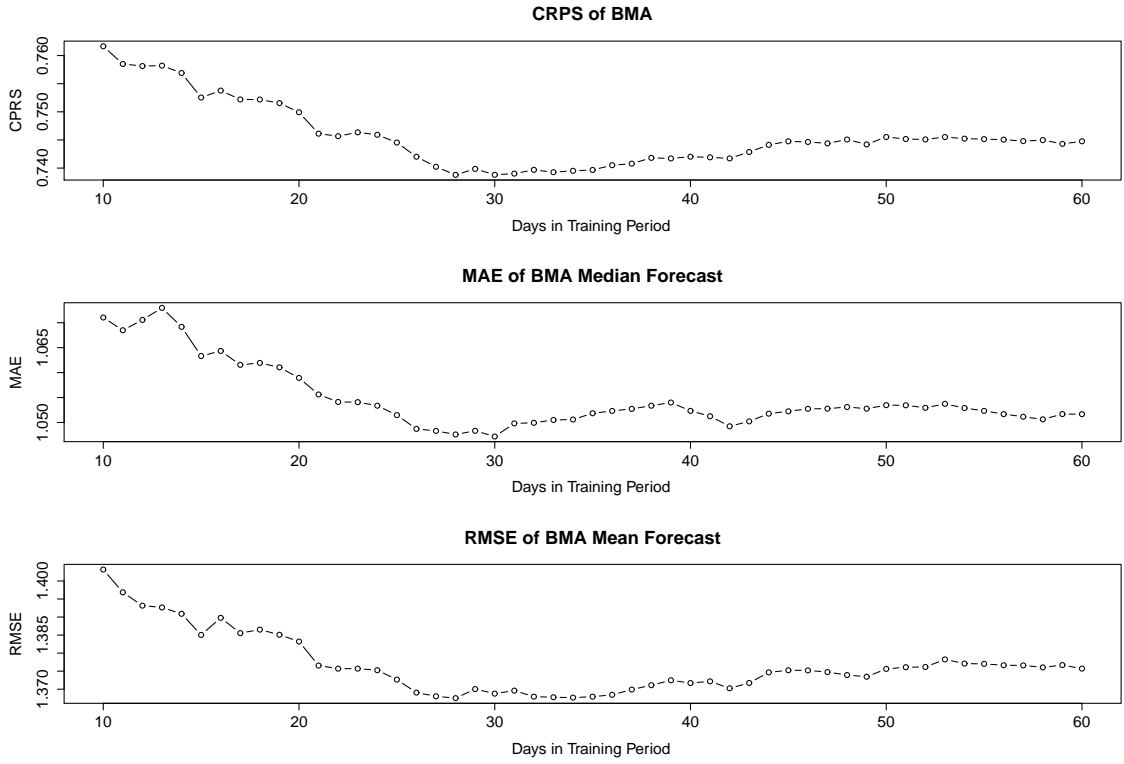,height=15cm, angle=-90}
\caption{CRPS of BMA predictive distribution, MAE values of BMA median and RMSE
  values of BMA mean forecasts
  corresponding to two-group model \eqref{eq:eq3.1} for various training period
  lengths.}   
\label{fig:fig5}
\end{center}
\end{figure}
Data analysis provided below was performed with the help of the {\tt
  ensembleBMA} package of R \citep{frgs,frgsb}. As a first step the
length of the 
appropriate training period was determined, then the performances
of the BMA post-processed ensemble forecasts corresponding to models
\eqref{eq:eq3.1} and \eqref{eq:eq3.2} were analyzed.  

\subsection{Training period}
   \label{sec:sub4.1}

According to the results of e.g. \citet{rgbp} to determine the length
of the training period to be used we compare the MAE values
of BMA median forecasts, the RMSE values of BMA mean forecasts, the
CRPS values of BMA predictive 
distributions and the coverages and average
widths of  $90\,\%$ and $66.7\,\%$ BMA prediction intervals for
training periods of 
length \ $10,11, \ldots, 60$ \ calendar days. In order to ensure
the comparability of 
the results we consider verification results from 02.12.2010 to
25.03.2011 (114 days).

\begin{figure}[t!]
\begin{center}
\leavevmode
\epsfig{file=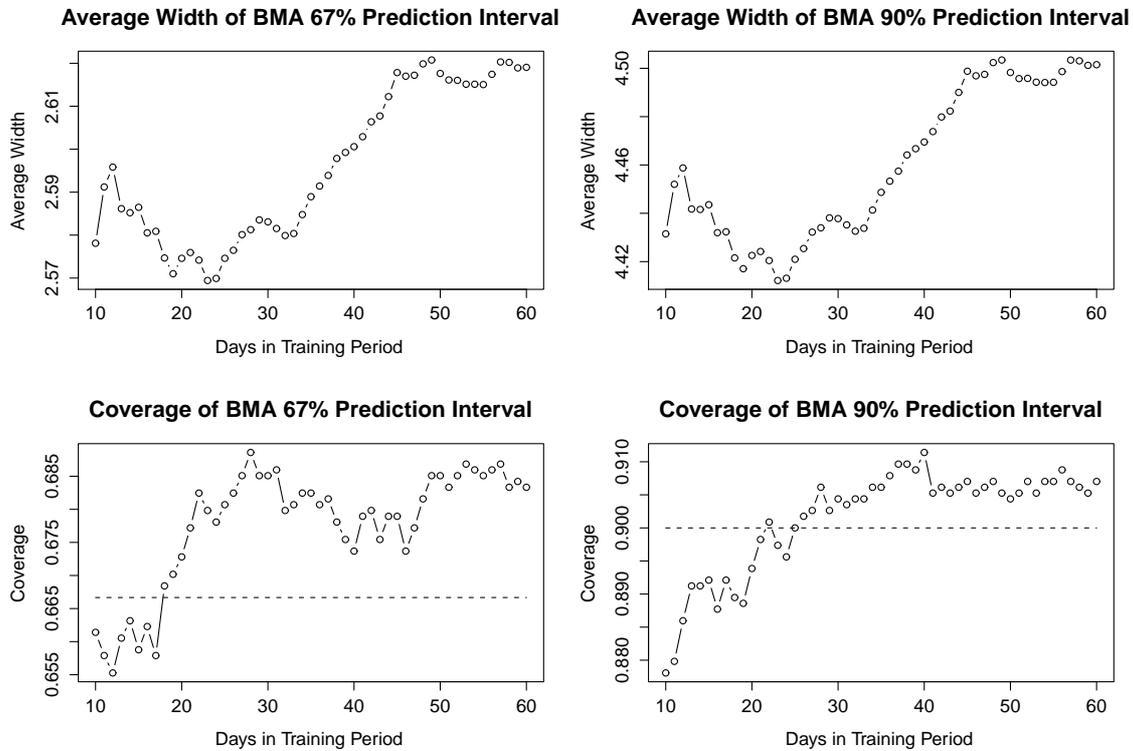,height=15cm, angle=-90}
\caption{Average widths and coverages of $66.7\,\%$ and $90\,\%$
 BMA  prediction intervals 
  corresponding to three-group model \eqref{eq:eq3.2} for various
  training period 
  lengths.}   
\label{fig:fig6}
\end{center}
\end{figure}
Consider first the two-group model \eqref{eq:eq3.1}. On Figure
\ref{fig:fig4} the 
average widths and coverages of $66.7\,\%$ and $90\,\%$  BMA prediction
intervals are plotted against the length of the training period. The
average widths of the prediction intervals  show an 
increasing trend, so shorter training periods yield sharper forecasts.
Coverages of $66.7\,\%$ and $90\,\%$ prediction intervals are not monotonously
increasing, too. For short training periods the coverage of the
$66.7\,\%$  prediction interval oscillates around the
correct $66.7\,\%$, but for training periods not shorter than 17 days it
stays above this level. The coverage of the $90\,\%$  prediction
interval stabilizes above the correct $90\,\%$ for training periods
longer than 24 days. Hence, to have calibrated forecasts, one should
choose a training period not shorter than 25 days.

\begin{figure}[t!]
\begin{center}
\leavevmode
\epsfig{file=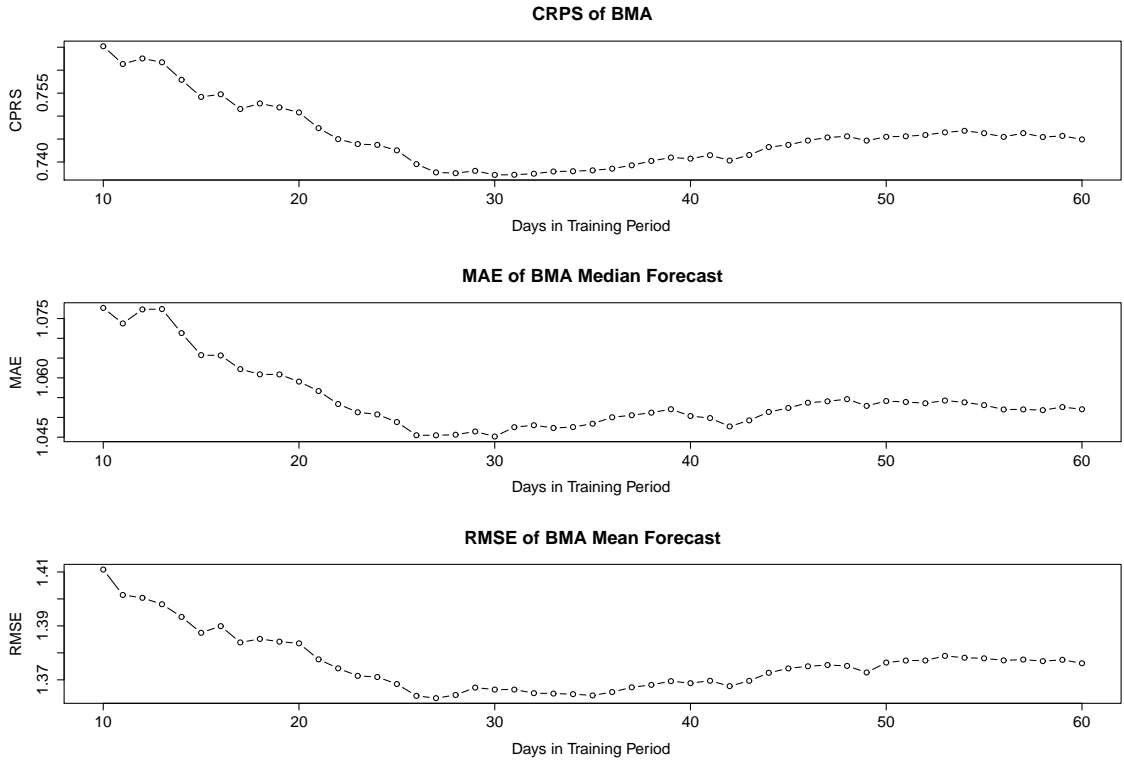,height=15cm, angle=-90}
\caption{CRPS of BMA predictive distribution, MAE values of BMA median and RMSE
  values of BMA mean forecasts
  corresponding to three-group model \eqref{eq:eq3.2} for various
  training period lengths.}   
\label{fig:fig7}
\end{center}
\end{figure}
Figure \ref{fig:fig5} shows
CRPS values of BMA predictive distribution, MAE values of BMA
median forecasts and RMSE values of BMA mean forecasts as functions of
the training period length. CRPS and RMSE both take their minima at 28
days, the corresponding values are $0.7388$ and $1.3675$,
respectively. MAE
takes its minimum of $1.0472$ at 30 days, while the second smallest value
($1.0476$) is obtained with a training period of length 28 days.
This means that for model
\eqref{eq:eq3.1} a 28 days training period seems to be reasonable and
training periods longer than 30 days cannot be taken into consideration.

Similar conclusions can be drawn from Figures \ref{fig:fig6} and
\ref{fig:fig7}  for the three-group model \eqref{eq:eq3.2}. In this case the
$66.7\,\%$ and $90\,\%$ prediction intervals are slightly narrower
than the corresponding intervals of model \eqref{eq:eq3.1}, their
coverages stabilize above the correct
$66.7\,\%$ and $90\,\%$ for training periods longer than 17 and 24
days, respectively. CRPS and MAE plotted on Figure \ref{fig:fig7} both
reach their minima of $0.7372$ and $1.0452$, respectively, at 30 days,
while values $0.7376$ and $1.0456$ corresponding to training period of
length 28 days are both the fourth smallest ones. RMSE takes its
minimum of $1.3632$ at 27 days, and increases afterwards. The fourth
smallest value ($1.3644$) again corresponds to 28 days, while the RMSE
corresponding to 30 days is significantly larger ($1.3664$). Moreover,
$66.7\,\%$ and $90\,\%$ prediction intervals corresponding to 28 days
are sharper than the appropriate prediction intervals calculated using training
period of length 30 days ($2.5813$ and $4.4340$ vs. $2.5831$ and
$4.4378$). Hence, we suggest the use of a training period of length 28
days for both BMA models.

\subsection{Predictions using BMA post-processing}
\label{sec:sub4.2}

According to the results of the previous subsection, to test the performance
of BMA post-processing on the 11 member ALADIN-HUNEPS ensemble we use a
training period of 28 calendar days. In this way ensemble members,
validating observations and BMA models are available for 146 calendar
days (on 20.11.2010 all ensemble members are missing). 

\begin{figure}[t!]
\begin{center}
\leavevmode
\epsfig{file=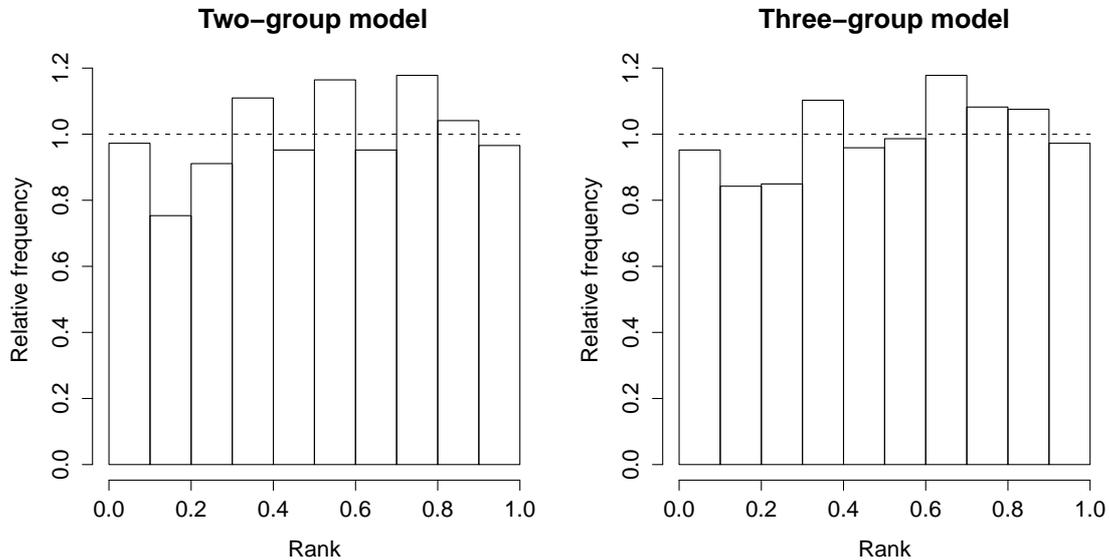,height=15cm, angle=-90}
\caption{PIT histograms for BMA post-processed forecasts using two-group
  \eqref{eq:eq3.1} and three-group \eqref{eq:eq3.2} models.}   
\label{fig:fig8}
\end{center}
\end{figure}

First we check the calibration of BMA post-processed forecasts with the
help of probability integral transform (PIT) histograms. The PIT is
the value of the BMA predictive cumulative distribution evaluated at
the verifying observations \citep{frg}. The closer the histogram to the
uniform distribution, the better is the calibration. On Figure
\ref{fig:fig8} the PIT histograms corresponding to two- and
three-group BMA models
\eqref{eq:eq3.1} and \eqref{eq:eq3.2} are displayed. Compared to the
verification rank histogram of the raw ensemble (see  Figure
\ref{fig:fig1}) one can observe a large improvement with the use of
calibration. However, these PIT histograms are still not perfect,
e.g. Kolmogorov-Smirnov test rejects uniformity both for the two- and
for the three-group model. The corresponding $p$-values are $0.0222$
and $0.0187$, respectively, so the PIT of the two-group model is
slightly better.

\begin{table}[b]
\begin{center}
\begin{tabular}{|l|c|c|c|c|c|} \hline
\multicolumn{1}{|c|}{}&
\multicolumn{2}{|c|}{Coverage ($\%$)}&
\multicolumn{2}{|c|}{Average Width}\\ \cline{2-5}
Interval&$66.7\,\%$ interval&$90.0\,\%$ interval&$66.7\,\%$
interval&$90.0\,\%$ interval\\ \hline 
Raw ensemble&$38.70$&$55.14$&$1.4388$&$2.2001$
\\ 
BMA model
\eqref{eq:eq3.1}&$68.08$&$90.34$&$2.6359$&$4.5297$
\\ 
BMA model
\eqref{eq:eq3.2}&$68.36$&$90.21$&$2.6153$&$4.4931$
\\ 
\hline 
\end{tabular} 
\caption{Coverage and average width of prediction intervals.} \label{tab:tab1}
\end{center}
\end{table}
Table \ref{tab:tab1} gives the coverages and average widths of
$66.7\,\%$ and $90.0\,\%$ prediction intervals calculated using models
\eqref{eq:eq3.1} and \eqref{eq:eq3.2}, and the corresponding measures calculated
from the raw ensembles. In the latter case the ensemble of forecasts
corresponding to a given location and time is considered as a
statistical sample. The BMA prediction intervals calculated from
both models are approximately twice as wide, as the corresponding
intervals of the raw ensemble. This comes from the small dispersion of
the raw ensemble, see the verification rank histogram of Figure
\ref{fig:fig1}. Concerning calibration one can observe that the
coverages of both  
BMA prediction intervals are rather close to the right coverages,
while the coverages of the prediction intervals calculated from the
raw ensemble are quite poor. This also shows that BMA post-processing
highly improves  calibration. Further, BMA model \eqref{eq:eq3.2}
yields slightly sharper predictions but there is no big difference
between the coverages of the two BMA models.

\begin{table}[t]
\begin{center}
\begin{tabular}{|l|c|c|c|c|c|c|c|c|c|c|} \hline
\multicolumn{1}{|c|}{}&
\multicolumn{1}{|c|}{Mean CRPS}&
\multicolumn{2}{|c|}{MAE}&
\multicolumn{2}{|c|}{RMSE}\\ \cline{3-6}
&&median&mean&median&mean
\\ \hline 
Raw ensemble&$0.8599$&$1.1215$&$1.1090$&$1.4634$&$1.4440$
\\ 
BMA model
\eqref{eq:eq3.1}&$0.7577$&$1.0678$&$1.0763$&$1.4213$&$1.4067$
\\ 
BMA model
\eqref{eq:eq3.2}&$0.7556$&$1.0643$&$1.0749$&$1.4153$&$1.4018$
\\\hline 
\end{tabular} 
\caption{Mean CRPS of probabilistic, MAE and RMSE of
deterministic forecasts.} \label{tab:tab2}
\end{center}
\end{table}

\begin{figure}[t]
\begin{center}
\leavevmode
\epsfig{file=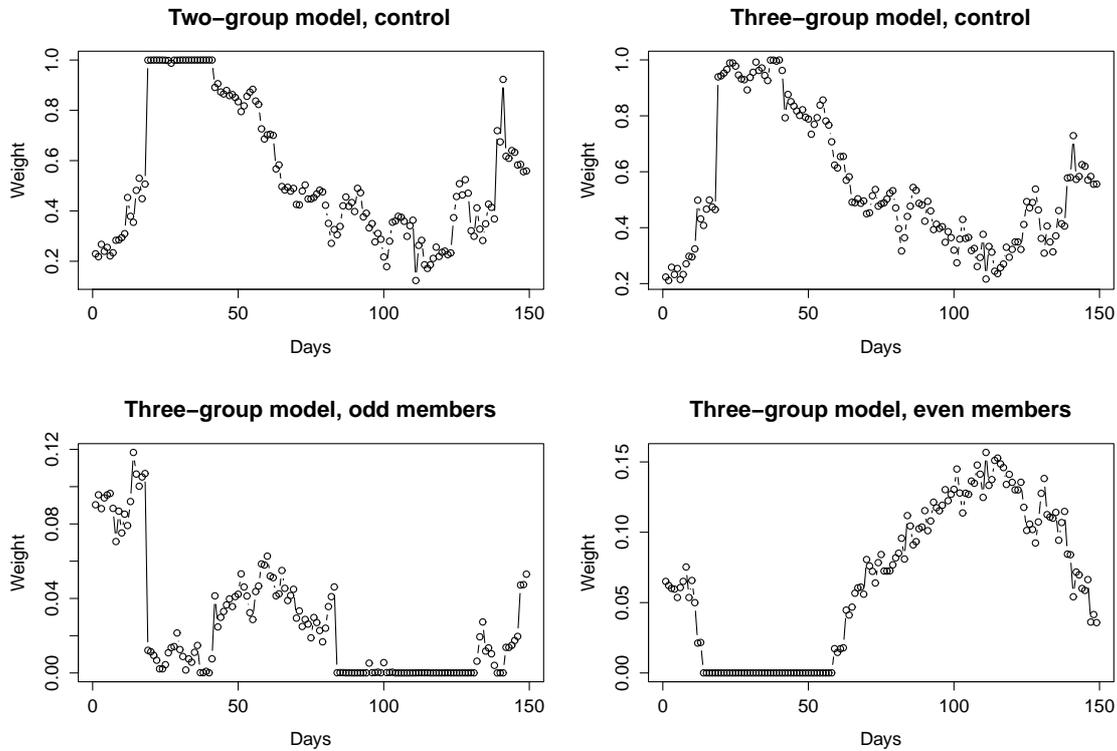,height=15cm, angle=-90}
\caption{BMA weights of two-group
  \eqref{eq:eq3.1} and three-group \eqref{eq:eq3.2} models.}   
\label{fig:fig9}
\end{center}
\end{figure}
On Table \ref{tab:tab2} the verification results of the model
fit are given. Verification measures of probabilistic forecasts and point
forecasts calculated using BMA models \eqref{eq:eq3.1} and
\eqref{eq:eq3.2} are compared to the corresponding measures calculated
for the raw ensemble. Examining these results one can clearly observe
the advantage of BMA post-processing which resulted a significant
decrease in all verification scores. Further, the BMA median forecasts
yield slightly lower MAE values than the BMA mean forecasts for both
models, while in the case of RMSE values the situation is just the
opposite, which is a perfect illustration of the theoretical results
of \cite{gneiting11} about the optimality of these verification scores. 
Finally, model  \eqref{eq:eq3.2} distinguishing three exchangeable
groups of ensemble forecasts slightly outperforms model \eqref{eq:eq3.1}.

Figure \ref{fig:fig9} shows the BMA weights corresponding to models
\eqref{eq:eq3.1} and \eqref{eq:eq3.2}. Examining the behaviour of 
weight \ $\omega$ 
\ of the control member of the ensemble in the two-group model
\eqref{eq:eq3.1}, one can 
observe that in $84.56\,\%$ of the cases there is a real mixture of
gamma distributions. The values of \ $\omega$ \ which are close to $1$
correspond to a
continuous time interval 17.11.2010 -- 09.12.2010,  when the control
member of the ensemble gives much
better forecasts than the ten exchangeable ensemble members. This can
clearly be seen from Table \ref{tab:tab3} where the MAE and RMSE
values of the particular ensemble members are given for the above
mentioned period. In all of these 23 subsequent days \ $\omega>0.995$ \
but on 20.11.2010, when \ $\omega=0.9873$. \ However, as it was
mentioned earlier, on this particular day all ensemble 
forecasts are missing from the data set. The situation is quite
different in the case of the three-group model \eqref{eq:eq3.2}, where
the weight \ $\omega_c$ \ of the control is close to $1$ (greater than
$0.98$) only on 7 days, so in the remaining cases ($95.30\,\%$)  a
real mixture of gamma distributions present. Further, observe that
there are 55 days \ ($36.91\,\%$) \ when all BMA weights
are positive, the even numbered exchangeable members have nearly zero weights \
(less than $0.001$) \ in 45 cases \ ($30.20\,\%$) \ at the beginning of the
considered time period, while the odd numbered exchangeable members
are almost zero in 53 cases \ ($35.57\,\%$), \ mainly at the end of
it.

\begin{table}[b]
\begin{center}
\begin{tabular}{|l|c|c|c|c|c|c|c|c|c|c|c|} \hline
\multicolumn{1}{|c|}{}&
\multicolumn{1}{|c|}{Control}&
\multicolumn{10}{|c|}{Exchangeable members}
\\ \cline{2-12}
&$f_c$&$f_{\ell,1}$&$f_{\ell,2}$&$f_{\ell,3}$&$f_{\ell,4}$&$f_{\ell,5}$&$f_{\ell,6}$&$
f_{\ell,7}$&$f_{\ell,8}$&$f_{\ell,9}$&$f_{\ell,10}$ \\ \hline
MAE&$1.32$&$1.60$&$1.46$&$1.52$&$1.68$&$1.51$&$1.49$&$1.56$&$
1.42$&$1.41$&$1.65$\\
RMSE&$1.69$&$2.16$&$1.86$&$1.96$&$2.26$&$1.92$&$1.95$&$2.05$&$
1.89$&$1.81$&$2.23$\\ \hline
\end{tabular} 
\caption{MAE and RMSE of the control and exchangeable ensemble
  forecasts for the period  17.11.2010 -- 09.12.2010.} \label{tab:tab3}
\end{center}
\end{table}

\begin{figure}[t]
\begin{center}
\leavevmode
\epsfig{file=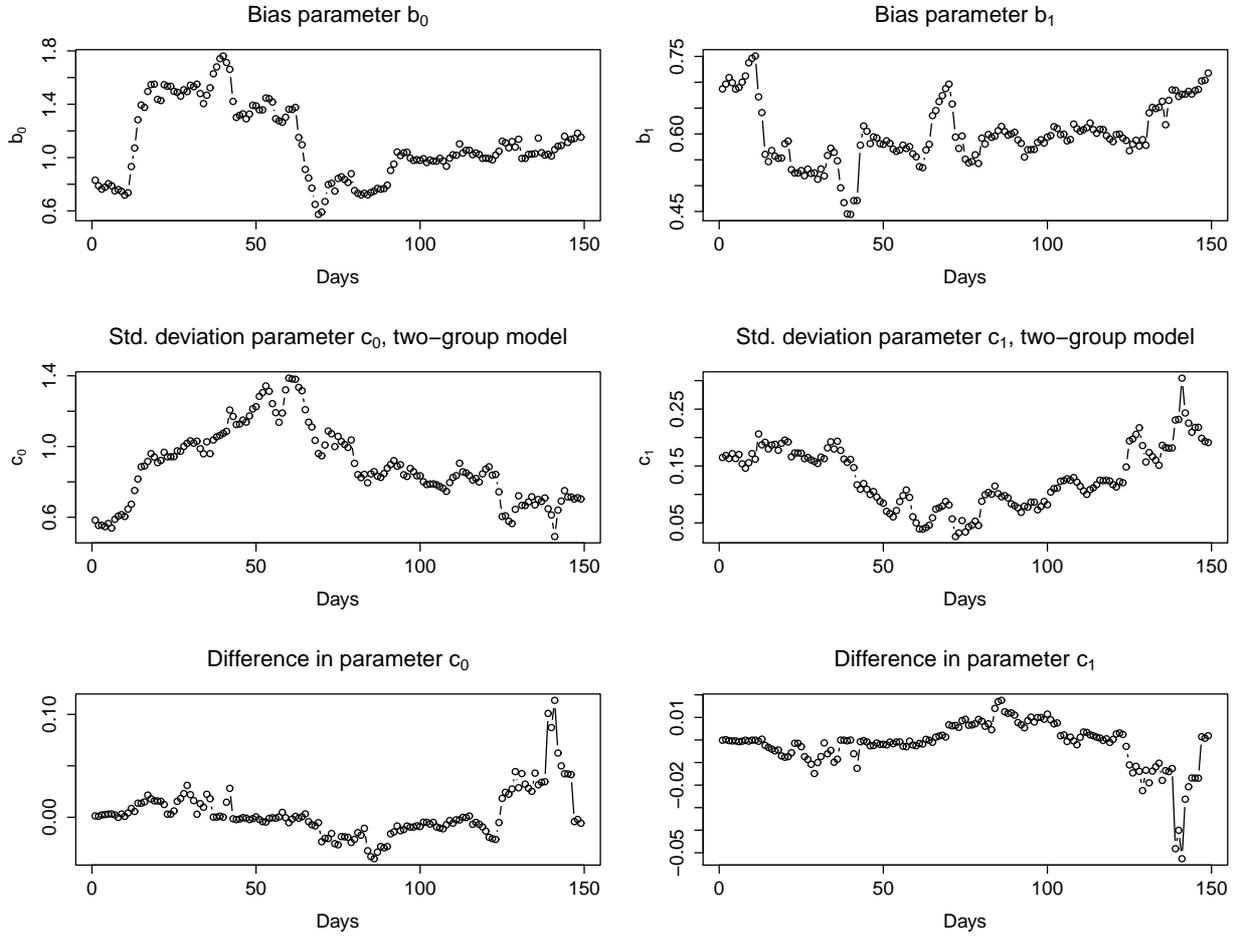,height=16.5cm, angle=-90}
\caption{Parameters of two-group model
  \eqref{eq:eq3.1} and differences in standard deviation parameters
  between three- and two-group models.}     
\label{fig:fig10}
\end{center}
\end{figure}
Finally, on Figure \ref{fig:fig10} common bias parameters \ $b_0, \ b_1$ \
of both BMA models investigated and standard deviation parameters \
$c_0, \ c_1$ \ 
of the two-group model \eqref{eq:eq3.1} are plotted, together with the
differences in standard deviation parameters of three- and two-group
models. Bias parameters are rather
stable, the relative standard deviations of \ $b_0$ \ and \ $b_1$ \
are $25.44\,\%$ and $9.97\,\%$, respectively. Hence, the BMA mean
forecast of a particular day is mainly determined by the corresponding
ensemble forecasts. The standard deviation 
parameters show more variability, for  \ $c_0$ \
and \ $c_1$ \ the relative standard deviations are equal to
$23.41\,\%$ and $41.27\,\%$ for 
model \eqref{eq:eq3.1}, and $22.64\,\%$ and $36.73\,\%$ for
model \eqref{eq:eq3.2}.

\section{Conclusions}
   \label{sec:sec5}

In the present study the BMA ensemble post-processing method is applied 
for the 11 member ALADIN-HUNEPS ensemble of the HMS to obtain 42 hour
predictions for 10 meter wind speed. Two different BMA models are
investigated, one assumes two groups of exchangeable members (control
and forecasts from perturbed initial conditions), while the other
considers three (control and forecasts from perturbed initial
conditions with positive and negative perturbations).  For both models
a 28 days training period is suggested. The comparison of the
raw ensemble and of the probabilistic forecasts shows that the mean
CRPS values 
of BMA post-processed forecasts are considerably lower than the mean
CRPS of the raw ensemble. Further, the MAE and RMSE values of BMA point
forecasts (median and mean) are also lower than the MAEs and RMSEs of the
ensemble median and of the ensemble mean. The calibrations of BMA
forecasts are nearly perfect, the coverages of $66.7\,\%$ and $90.0\,\%$
prediction intervals are very close to the right values. The
three-group BMA model slightly outperforms the two-group one and in
almost all cases yields a real mixture of gamma distributions.

In this way one can conclude that BMA post-processing of ensemble
forecasts of wind speed data of the HMS significantly improves the
precision and calibration of the forecasts, its operational application
is worth considering.

\bigskip
\noindent
{\bf Acknowledgments.} \  \ Research was supported by 
the Hungarian  Scientific Research Fund under Grants No. OTKA
T079128/2009 and OTKA NK101680 and by the T\'AMOP-4.2.2.C-11/1/KONV-2012-0001
project. The project has been supported by the European Union,
co-financed by the European Social Fund.  The authors are indebted to Tilmann
Gneiting for his useful suggestions and remarks and to M\'at\'e Mile and 
Mih\'aly Sz\H ucs from the HMS for providing the data.

\end{document}